\documentclass[nofootinbib,10pt,twocolumn,pra,showpacs,tightenlines, showkeys, english,aps]{revtex4}
\usepackage[T1]{fontenc}
\usepackage{color}
\usepackage{array}
\usepackage{amsmath}
\usepackage{graphicx}

\usepackage{epsfig,graphicx,times}
\usepackage{amstext}
\usepackage{amsmath}            
\usepackage{amssymb}            
\usepackage{graphicx}           
\usepackage{latexsym}
\usepackage{bm}
\usepackage{color}
\usepackage{amssymb}
\usepackage{subfig}
\usepackage{epstopdf}
\usepackage{epsfig}
\usepackage{caption}
\makeatletter

\providecommand{\tabularnewline}{\\}

\usepackage{epsfig}\usepackage{bm}\usepackage{epsfig}\@ifundefined{definecolor}
 {\usepackage{color}}{}

\def\be{\begin{equation}}
\def\ee{\end{equation}}
\def\bea{\begin{eqnarray}}
\def\eea{\end{eqnarray}}

\@ifundefined{showcaptionsetup}{}{%
 \PassOptionsToPackage{caption=false}{subfig}}
\usepackage{subfig}
\makeatother

\usepackage{babel}

\begin{document}

\title{Protocols and quantum circuits for implementing entanglement concentration
in cat state, GHZ-like state and 9 families of 4-qubit entangled states}

\author{Chitra Shukla$^{a}$, Anindita Banerjee$^{b}$, Anirban Pathak$^{a,c,}$%
\footnote{Email: anirban.pathak@gmail.com,  Phone: +91 9717066494}}

\affiliation{
$^{a}$Jaypee Institute of Information Technology, A-10, Sector-62,
Noida, UP-201307, India\\
${b}$ Department of Physics and Center for Astroparticle Physics
and Space Science, Bose Institute, Block EN, Sector V, Kolkata 700091,
India\\
$^{c}$RCPTM, Joint Laboratory of Optics of Palacky University and
Institute of Physics of Academy of Science of the Czech Republic,
Faculty of Science, Palacky University, 17. listopadu 12, 771 46 Olomouc,
Czech Republic}

\begin{abstract}
Three entanglement concentration protocols (ECPs) are proposed. The
first ECP and a modified version of that are shown to be useful for
the creation of maximally entangled cat and GHZ-like states from their
non-maximally entangled counterparts. The last two ECPs are designed
for the creation of maximally entangled $(n+1)$-qubit state $\frac{1}{\sqrt{2}}\left(|\Psi_{0}\rangle|0\rangle+|\Psi_{1}\rangle|1\rangle\right)$
from the partially entangled $(n+1)$-qubit normalized state $\alpha|\Psi_{0}\rangle|0\rangle+\beta|\Psi_{1}\rangle|1\rangle,$
where $\langle\Psi_{1}|\Psi_{0}\rangle=0$ and $|\alpha|\neq\frac{1}{\sqrt{2}}$.
It is also shown that W, GHZ, GHZ-like, Bell and cat states and specific
states from the 9 SLOCC-nonequivalent families of 4-qubit entangled
states can be expressed as $\frac{1}{\sqrt{2}}\left(|\Psi_{0}\rangle|0\rangle+|\Psi_{1}\rangle|1\rangle\right)$
and consequently the last two ECPs proposed here are applicable to
all these states. Quantum circuits for implementation of the proposed
ECPs are provided and it is shown that the proposed ECPs can be realized
using linear optics. Efficiency of the ECPs are studied using a recently
introduced quantitative measure (Phys. Rev. A \textbf{85},  012307 (2012)).
Limitations of the measure are also reported.

\end{abstract}
\maketitle

\section{Introduction\label{sec:Introduction}}

Entanglement plays a crucial role in quantum computation and quantum
communication \cite{Non local CNOT,Manu-annalender,Bennett-teleportation,Bennett-densecoding,ekert,Ping-Pong,DLL,Beyond GV}.
It is essential for realization of various protocols and algorithms
\cite{Non local CNOT,Manu-annalender,Bennett-teleportation,Bennett-densecoding,ekert,Ping-Pong,DLL,Beyond GV}.
Specifically, maximally entangled state is required for realization
of nonlocal quantum gates \cite{Non local CNOT}, distributed quantum
search algorithm \cite{Manu-annalender}, teleportation \cite{Bennett-teleportation},
densecoding \cite{Bennett-densecoding}, entanglement based quantum
key distribution \cite{ekert} and entanglement based secure direct
quantum communication \cite{Ping-Pong,DLL,Beyond GV}. In these applications,
entanglement is generally produced locally and distributed to different
parties involved in the communication or computation process. During
the transmission, processing and storing a maximally entangled pure
state may interact with the environment and become mixed state or
less entangled (i.e., non-maximally entangled) pure state. This may
happen because of many different reasons. For example, the transmission
channel may be noisy. In general, the amount of entanglement is usually
reduced during transmission process. Unfortunately, such degradation
of entanglement is unavoidable, but for the proper execution of the
protocols mentioned above \cite{Manu-annalender,Bennett-teleportation,Bennett-densecoding,ekert,Ping-Pong,DLL,Beyond GV}
we need perfect quantum channels for distribution of ebits. In absence
of such a channel it would be sufficient to design a protocol that
can convert the non-maximally entangled state back into maximally
entangled state. Interestingly, such protocols exist and the protocols
are divided into two classes (i) Entanglement concentration protocol
(ECP) which can transform a partially entangled pure state (pure non-maximally
entangled state) into a maximally entangled state and (ii) Entanglement
purification/distillation protocols (EP) that can transform a mixed
non-maximally entangled state into a maximally entangled state. In
1996, Bennet et al. \cite{origional ECP} proposed the first ECP.
In the same year they proposed an EP, too \cite{Entaglement purification}.
In their pioneering work Bennet et al. used collective entanglement
concentration procedure (i.e., Schmidt projective method). Since the
pioneering works of Bennet et al., several ECPs and EPs are proposed
\cite{S. Bose ,bandyopadhyay,zhao-proposal,yamamoto,Sheng-1,entropy,Bell-determinstic-ckt,Bell_Deng,Sheng-2,opt-comm-2013-he,Dhara GHZ state,GHZ-zhou,Dhara Cluster state,T Xu-cluster,Zhao-cluster,NOON}.
Initially most of the ECPs were proposed for non-maximally entangled
Bell state. Specifically, Bose et al. \cite{S. Bose }, Bandyopadhyay
\cite{bandyopadhyay}, Zhao et al. \cite{zhao-proposal}, Yamamoto
et al. \cite{yamamoto}, Sheng et al. \cite{Sheng-1}, Sheng and Zhou
\cite{entropy}, Gu et al. \cite{Bell-determinstic-ckt}, Deng \cite{Bell_Deng}
proposed ECPs and EPs for partially entangled Bell state. While Bose
et al.'s proposal involved entanglement swapping and Bell state measurement
and Gu et al.'s proposal involved projective operator valued measurement
(POVM), other proposals \cite{zhao-proposal,yamamoto,Sheng-1} circumvented
the use of Bell measurement and POVM and discussed the possibilities
of optical implementation of proposed ECPs using polarizing beam splitter
(PBS) and wave-plates. In fact, ECPs proposed by Zhao et al. and Yamamoto
et al. were experimentally realized in 2003 \cite{experiment2-zhao,experiment1-yamamoto}.
It was shown in some of these initial works that the ECPs designed
for partially entangled Bell states can be generalized to build ECPs
for partially entangled cat ($N$-partite GHZ) states. Recently, several
efforts have been made to extend the applicability of ECPs beyond
the production of Bell states. For example, in 2012, Sheng et al.
have proposed an ECP for partially entangled arbitrary W state \cite{Sheng-2}.
In 2013, Ling-yan He \cite{opt-comm-2013-he} proposed a single nitrogen
vacancy (N-V) center assisted ECP for a specific type of partially
entangled W state. Very recently, ECPs for partially entangled GHZ
states are proposed by Choudhury and Dhara \cite{Dhara GHZ state}
and Zhou et al. \cite{GHZ-zhou}, ECPs for 4-qubit cluster state are
proposed by Choudhury and Dhara \cite{Dhara Cluster state}, Ting-Ting
Xu et al. \cite{T Xu-cluster} and Zhau et al. \cite{Zhao-cluster},
ECP for partially entangled NOON states is proposed by Zhou et al.
\cite{NOON}. Clearly, much attention has recently been paid to develop
ECPs for partially entangled states other than Bell state. Interestingly,
majority of these recent efforts are concentrated toward the construction
of ECPs for partially entangled GHZ states {[}\cite{Dhara GHZ state,GHZ-zhou}
and references therein{]} and W states {[}\cite{Sheng-2,opt-comm-2013-he}
and references therein{]}. In case of 3-qubit pure states it is well-known
that there are only 2 families of entangled states \cite{3-qubit SLOCC}
under stochastic local quantum operations assisted by classical communication
(SLOCC). These two families of 3-qubit entangled states are referred
to as GHZ and W states. Thus the recent trend of developing ECPs for
partially entangled GHZ and W states is reasonable. Interestingly,
no ECP has yet been proposed to explicitly concentrate GHZ-like states.
However, in recent past several applications of maximally entangled
GHZ-like states have been reported \cite{GHZ-like-1,GHZ-like2,GHZ-like3}.
Keeping this in mind, in the present work we have proposed an ECP
for partially entangled cat state and have modified that to develop
an ECP for partially entangled GHZ-like state. 

Recent success in developing ECPs for both the families of 3-qubit entangled states (i.e., for GHZ and W class) also motivated
us to ask: How to concentrate pure states of different families of
4-qubit partially entangled states? Present paper's main objective
is to answer this question. The effort is timely as to the best of
our knowledge until date for 4-qubit pure states ECPs are proposed
only for partially entangled 4-qubit cat state and 4-qubit cluster
state. No effort has yet been made to concentrate other families of
4-qubit entangled states. Keeping this in mind, present paper is focused
around construction of a general ECP for 9 families of 4-qubit entangled
states. Before we discuss further detail of our idea, it would be
apt to note that in 2002, Verstraete et al. \cite{4-qubit SLOCC}
had shown that 4-qubit pure states can be entangled in 9 different
ways under SLOCC. A rigorous proof of this classification of 4-qubit
pure states was subsequently provided by Chterental and Djokovic in
2007 \cite{4-Qubit-SLOCC-proof}. In 2010, a similar SLOCC classification
of 4-qubit pure states was obtained using string theory \cite{string}.
However, in 2010, another interesting result was reported by Gaur
and Wallach \cite{gilad}, in which they had established the existence
of uncountable number SLOCC-nonequivalent classes of 4-qubit entangled
states. In what follows we will use Verstraete et al.'s \cite{4-qubit SLOCC}
classification and propose two ECPs that can concentrate some states
from each of the 9 families proposed by Verstraete et al. \cite{4-qubit SLOCC}. 

Until now different strategies have been used for developing ECPs
and EPs. For example, ECPs are proposed using linear optics (specifically,
using PBSs and wave plates), cross-Kerr-nonlinearities (i.e., using
quantum non-demolition (QND) measurements) \cite{Sheng-1,Sheng-2},
entanglement swapping and Bell measurement \cite{S. Bose }, unitary
transformation \cite{zhao-proposal} and quantum electrodynamics (QED)
based techniques \cite{QED} etc. However, most of the recent works
discuss ECP with a perspective where qubits are realized using the
polarization of photon. As the qubit can be realized using different
systems, such as superconductivity, NMR, photon etc., in what follows
we have not restricted ourselves to any specific technology and have
presented our protocol in general as a quantum circuit. This makes
it applicable to any specific kind of realization of qubits. Only
at the end of the paper we have shown that the present work can be
realized using PBSs, wave-plates and photon-detectors. As the proposed
protocols do not require anything other than implementation of Bell
measurement it can also be realized in other implementations of qubits.
For example, it is straight forward to implement the ECPs proposed
here using NMR-based approach as Bell measurement is possible in NMR
\cite{NMR-Anil Kumar}. 

Rest of the paper is organized as follows. In Section \ref{sec:Entanglement-concentration-protocols}
an ECP for cat state is proposed and it is modified to develop an
ECP for GHZ-like state. Quantum circuits for these ECPs are also described.
In Section \ref{sec:A-generalized-scheme} we propose two ECPs for
$(n+1)$-qubit normalized states of the form $|\psi\rangle=\alpha|\Psi_{0}\rangle|0\rangle+\beta|\Psi_{1}\rangle|1\rangle,$
where $|\Psi_{0}\rangle$ and $|\Psi_{1}\rangle$ are mutually orthogonal
$n$-qubit states and $|\alpha|\neq\frac{1}{\sqrt{2}}$. In this section
we have also shown that each of the 9 families of four qubit states
contain some states of the form $\alpha|\Psi_{0}\rangle|0\rangle+\beta|\Psi_{1}\rangle|1\rangle.$
In Section \ref{sec:Optical-implementation} we have shown that the
ECPs proposed here can be realized using linear optics. In Section
\ref{sec:Efficiency} efficiencies of the proposed ECPs are discussed
using a quantitative measure of ECP introduced by Sheng et al. \cite{Sheng-1}
and finally the paper is concluded in Section \ref{sec:Conclusion}.

\section{Entanglement concentration protocols (ECPs) \label{sec:Entanglement-concentration-protocols}}

In the previous section we have already described the importance of
ECPs in quantum information processing. In this section we propose
new ECPs for partially entangled cat state and GHZ-like states. To
begin with we first propose an ECP for a non-maximally entangled cat
state.

\subsection{ECP for partially entangled cat state\label{sub:For-Cat-state}}

\begin{figure}
\centering{}\includegraphics[scale=0.7]{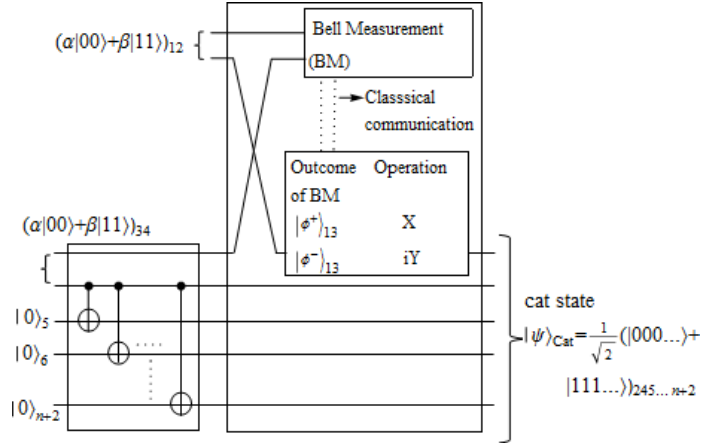}\caption{\label{fig:Quantum-circuit-for-cat}Quantum circuit for ECP of non-maximally
entangled cat state $|\psi\rangle_{{\rm cat}}$ }
\end{figure}

A non-maximally entangled Bell-type state may be defined as 

\begin{equation}
|\psi\rangle_{{\rm Bell}}=(\alpha|00\rangle+\beta|11\rangle)_{12},\label{eq:Bell state (2 qubit)}\end{equation}
where $|\alpha|^{2}+|\beta|^{2}=1$ and $|\alpha|\neq|\beta|.$ Similarly,
we may define a non-maximally entangled $n$-qubit cat state as 

\begin{equation}
|\psi\rangle_{{\rm cat}}=(\alpha|000\cdots0\rangle+\beta|111\cdots1\rangle)_{12\cdots n}.\label{eq:cat state}\end{equation}
We wish to devise an ECP for $|\psi\rangle_{{\rm cat}}$ with the
help of non-maximally entangled Bell state $|\psi\rangle_{{\rm Bell}}.$
For this purpose we introduce the quantum circuit shown in Fig. \ref{fig:Quantum-circuit-for-cat}.
The circuit is composed of two parts. In the first part (See the left
most box of Fig. \ref{fig:Quantum-circuit-for-cat}) we produce non-maximally
entangled $n$-qubit cat state $|\psi\rangle_{{\rm cat}}$ starting
from a non-maximally entangled Bell state $|\psi\rangle_{{\rm Bell}}$
and $(n-2)$ auxiliary qubits each prepared in $|0\rangle$. Working
of this part of the circuit may be understood as follows: Assume that
we have a non-maximally entangled $n$-qubit cat state $|\psi\rangle_{{\rm cat}}$
and we add an auxiliary qubit (prepared in $|0\rangle$) with that
as $(n+1)^{th}$ qubit and apply a ${\rm CNOT}$ operation with any
one of the qubits of the $n$-qubit $|\psi\rangle_{{\rm cat}}$ state
(in Fig. \ref{fig:Quantum-circuit-for-cat} it is the second qubit
of $n$-qubit $|\psi\rangle_{{\rm cat}}$) as the control qubit and
the auxiliary qubit as the target qubit. This would yield an $(n+1)$-qubit
$|\psi\rangle_{{\rm cat}}$ state as 

\begin{widetext}

 \begin{equation}
\begin{array}{lcl}
{\rm CNOT_{2\rightarrow n+1}}(|\psi\rangle_{{\rm cat}}\otimes|0\rangle) & = & {\rm CNOT_{2\rightarrow n+1}}((\alpha|000\cdots0\rangle+\beta|111\cdots1\rangle)_{12\cdots n}\otimes|0\rangle_{n+1})\\
 & = & {\rm CNOT_{2\rightarrow n+1}}(\alpha|000\cdots00\rangle+\beta|111\cdots10\rangle)_{12\cdots n+1}\\
 & = & (\alpha|000\cdots00\rangle+\beta|111\cdots11\rangle)_{12\cdots n+1}.\end{array}\label{eq:CNOT CAT state}\end{equation}
\end{widetext}
This part of the circuit is not the main component of the proposed
ECP and it can be ignored. However, this may be relevant in the following
scenario: Assume that we have a machine for generation of maximally
entangled Bell state $|\psi^{+}\rangle=\frac{|00\rangle+|11\rangle}{\sqrt{2}}$.
The machine is not working properly and producing $|\psi\rangle_{{\rm Bell}}$
(with a fix but unknown value of $\alpha$) instead of $|\psi^{+}\rangle$.
In such a scenario we may first use an output of that imperfect Bell
state generator and $(n-2)$ auxiliary qubits prepared in $|0\rangle$
to produce $|\psi\rangle_{{\rm cat}}$. Once we obtain $|\psi\rangle_{{\rm cat}}$
(either prepared from imperfect Bell state generator or supplied)
we may use an entanglement swapping operation between another output
of that machine $\left(|\psi\rangle_{{\rm Bell}}\right)$ and $|\psi\rangle_{{\rm cat}}$
to obtain the desired ECP for $|\psi\rangle_{{\rm cat}}$ as shown
in the right most box of the circuit shown in Fig. \ref{fig:Quantum-circuit-for-cat}.
This part of the circuit works as follows: As the input of the main
part of the circuit for implementation of the ECP is $|\psi_{1}\rangle=|\psi\rangle_{{\rm Bell}}\otimes|\psi\rangle_{{\rm cat}}$
using (\ref{eq:Bell state (2 qubit)}) and (\ref{eq:CNOT CAT state})
we can write the input state as
\begin{widetext}
\begin{equation}
\begin{array}{lcl}
|\psi_{1}\rangle & = & (\alpha|00\rangle+\beta|11\rangle)_{12}\otimes(\alpha|000\cdots0\rangle+\beta|111\cdots1\rangle)_{345\cdots n+2}\\
 & = & (\alpha^{2}|00000\cdots0\rangle+\alpha\beta|00111\cdots1\rangle+\alpha\beta|11000\cdots0\rangle+\beta^{2}|11111\cdots1\rangle)_{12345\cdots n+2}.\end{array}\label{eq:Intermidiate Cat state 1}\end{equation}

Now after applying the SWAP gate shown in the circuit (i.e., after
swapping the second and third qubits of $|\psi\rangle_{{\rm 1}}$
we obtain

\begin{equation}
\begin{array}{lcl}
|\psi_{2}\rangle & = & \frac{1}{\sqrt{2}}[(\alpha^{2}(|\psi^{+}\rangle+|\psi^{-}\rangle)_{13}|000\cdots0\rangle_{245\cdots n+2}+\alpha\beta(|\phi^{+}\rangle+|\phi^{-}\rangle)_{13}|011\cdots1\rangle_{245\cdots n+2}\\
 & + & \alpha\beta(|\phi^{+}\rangle-|\phi^{-}\rangle)_{13}|100\cdots0\rangle_{245\cdots n+2}+\beta^{2}(|\psi^{+}\rangle-|\psi^{-}\rangle)_{13}|111\cdots1\rangle)_{245\cdots n+2}]\\
 & = & |\psi^{+}\rangle_{13}\frac{(\alpha^{2}|000\cdots0\rangle+\beta^{2}|111\cdots1\rangle)_{245\cdots n+2}}{\sqrt{2}}+|\psi^{-}\rangle_{13}\frac{(\alpha^{2}|000\cdots0\rangle-\beta^{2}|111\cdots1\rangle)_{245\cdots n+2}}{\sqrt{2}}\\
 & + & |\phi^{+}\rangle_{13}\frac{\alpha\beta(|011\cdots1\rangle+|100\cdots0\rangle)_{245\cdots n+2}}{\sqrt{2}}+|\phi^{-}\rangle_{13}\frac{\alpha\beta(|011\cdots1\rangle-|100\cdots0\rangle)_{245\cdots n+2}}{\sqrt{2}},\end{array}\label{eq: cat state final}\end{equation}
where we have used $|\psi^{\pm}\rangle=\frac{|00\rangle\pm|11\rangle}{\sqrt{2}}$
and $|\phi^{\pm}\rangle=\frac{|01\rangle\pm|10\rangle}{\sqrt{2}}$.
\end{widetext}

\begin{figure}
\centering{}\includegraphics[scale=0.7]{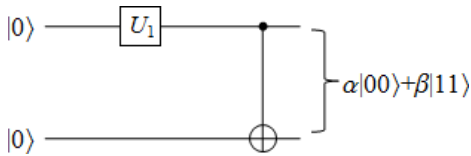}\caption{\label{fig:fig 2}Quantum circuit for generation of $|\psi\rangle_{{\rm Bell}}=\alpha|00\rangle+\beta|11\rangle$
where $U_{1}=\left(\protect\begin{array}{lc}
\alpha & -\beta^{*}\protect\\
\beta & \alpha^{*}\protect\end{array}\right)$}
\end{figure}

After the application of the SWAP gate a Bell measurement is performed
on the first two qubits of $|\psi\rangle_{2}.$ As a consequence of
that remaining $n$-qubits would collapse to one of the four states
as shown in Column 2 of Table \ref{tab:tab1}. If the outcome of the
Bell measurement is $|\psi^{\pm}\rangle$ then the protocol fails,
otherwise depending upon the outcome we apply a single qubit unitary
operation as shown in Column 3 of Table \ref{tab:tab1}. It is easy
to observe that on application of the single qubit unitary operation
in both the cases (i.e., if the outcome of Bell measurement is $|\phi^{+}\rangle$
or $|\phi^{-}\rangle$) we obtain a maximally entangled $n$-qubit
cat state $\frac{|000\cdots0\rangle+|111\cdots1\rangle}{\sqrt{2}}.$
Thus the circuit shown in Fig. \ref{fig:Quantum-circuit-for-cat}
is equivalent to an ECP for non-maximally entangled cat state. Knowledge
of values of $\alpha$ and $\beta$ are not required in the above
scenario where we create the state $|\psi\rangle_{{\rm cat}}$ starting
from a defective Bell state generator which always produces $|\psi\rangle_{{\rm Bell}}$
with same values of $\alpha$ and $\beta$. In case of any other scenario,
we would require to know the value of $\alpha$ or $\beta$ and use
that to produce $|\psi\rangle_{{\rm Bell}}$. Design of such a circuit
is a trivial exercise as $|\psi\rangle_{{\rm Bell}}$ may be easily
created using a modified EPR circuit where the Hadamard gate is replaced
by a single qubit unitary gate which maps $|0\rangle\rightarrow\alpha|0\rangle+\beta|1\rangle.$
A specific example of such a circuit is shown in Fig. \ref{fig:fig 2}
where the single qubit gate $U_{1}$ that maps $|0\rangle\rightarrow\alpha|0\rangle+\beta|1\rangle$
is introduced as \begin{equation}
U_{1}=\left(\begin{array}{lc}
\alpha & -\beta^{*}\\
\beta & \alpha^{*}\end{array}\right)\label{eq:u1}\end{equation}
In general, the proposed ECP is probabilistic. The existing ECPs contain
the same probabilistic nature but the feature is not explicitly mentioned.
For example, ECPs proposed in recent works of Choudhury and Dhara
\cite{Dhara GHZ state,Dhara Cluster state} are essentially probabilistic.
However, they didn't mention the probabilistic nature. To be consistent
with the conventional ECPs, we may assume that Alice prepares an $n$-qubit
cat state, keeps the first qubit with herself and sends the remaining
$n-1$ qubits to $n-1$ parties, say ${\rm Bob_{1},\, Bob_{2},\,\cdots,Bob_{n-1}.}$
At a later time the transmitted cat state may be reduced to partially
entangled cat state $|\psi\rangle_{{\rm cat}}.$ In order to concentrate
that in the above described protocol Alice prepares $|\psi\rangle_{{\rm Bell}},$
swaps her share of the $|\psi\rangle_{{\rm cat}}$ with the second
qubit of $|\psi\rangle_{{\rm Bell}},$ performs a Bell measurement
on the first two qubits of her possession and if the protocol succeeds
then she applies appropriate unitary operation (as described in Table
\ref{tab:tab1}) on the third qubit to obtain a maximally entangled
cat state shared between her and ${\rm Bob_{1},\, Bob_{2},\,\cdots,Bob_{n-1}.}$
ECPs proposed in the remaining part of this paper can also be illustrated
in the similar fashion. However, protocols proposed in the remaining
part of the paper are not described in this manner as it is a trivial
exercise.

\begin{widetext}
\begin{table}
\begin{centering}
\begin{tabular}{|c|c|c|c|}
\hline 
Outcome of Bell measurement &  State of the remaining $n$ qubits & Operation applied on qubit 2 & Final state\tabularnewline
\hline
$|\psi^{+}\rangle_{13}$ & $\frac{(\alpha^{2}|000\cdots0\rangle+\beta^{2}|111\cdots1\rangle)_{245\cdots n+2}}{\sqrt{2}}$ & \multicolumn{2}{c|}{Protocol fails}\tabularnewline
\hline
$|\psi^{-}\rangle_{13}$ & $\frac{(\alpha^{2}|000\cdots0\rangle-\beta^{2}|111\cdots1\rangle)_{245\cdots n+2}}{\sqrt{2}}$ & \multicolumn{2}{c|}{Protocol fails}\tabularnewline
\hline 
$|\phi^{+}\rangle_{13}$ & $\frac{\alpha\beta(|011\cdots1\rangle+|100\cdots0\rangle)_{245\cdots n+2}}{\sqrt{2}}$ & $X$ & $\frac{|000\cdots0\rangle+|111\cdots1\rangle}{\sqrt{2}}$\tabularnewline
\hline 
$|\phi^{-}\rangle_{13}$ & $\frac{\alpha\beta(|011\cdots1\rangle-|100\cdots0\rangle)_{245\cdots n+2}}{\sqrt{2}}$ & $iY$ & $\frac{|000\cdots0\rangle+|111\cdots1\rangle}{\sqrt{2}}$\tabularnewline
\hline
\end{tabular}
\par\end{centering}

\caption{\label{tab:tab1}Relation among Alice's Bell state measurement outcome,
cat state and operation applied }
\end{table}
\end{widetext}

\subsubsection{Special cases of cat state}

Interestingly Bell state and GHZ state are special cases of an $n$-qubit
cat state for $n=2$ and $n=3,$ respectively. Thus the Bell-measurement
based ECP presented here also works as ECP for non-maximally entangled
Bell and GHZ states. In 1999, S. Bose et al. \cite{S. Bose } proposed
a Bell-measurement based scheme for obtaining the maximally entangled
Bell state $|\psi^{+}\rangle$ from $|\psi\rangle_{{\rm Bell}}$.
Bose et al.'s scheme can now be viewed as a special case of the above
proposed ECP for cat state for $n=2$. Similarly, for $n=3$ qubits
proposed ECP reduces to an ECP for GHZ state. Very recently a Bell-measurement
based ECP for a non-maximally entangled 3-qubit GHZ state is proposed
by Choudhury and Dhara \cite{Dhara GHZ state}. Their ECP can also
be viewed as a special case of our ECP for cat state. The scheme proposed
here is more general than Choudhury and Dhara \cite{Dhara GHZ state}
scheme for several other reasons, too. For example, $\alpha$ and
$\beta$ are unknown and complex numbers here while in work of Choudhury
and Dhara $\alpha$ and $\beta$ were considered as real and known.
Further, complexity of their approach is high as the ECP proposed
by Choudhury and Dhara involves many steps that are not essential.

\subsection{ECP for partially entangled GHZ-like state\label{sub:For-GHZ-like-state}}

\begin{figure}
\centering{}\includegraphics[scale=0.7]{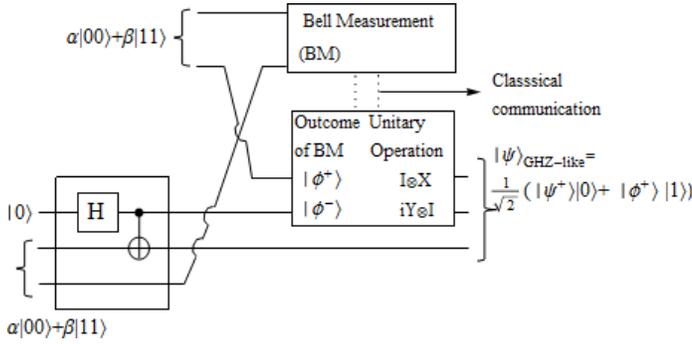}\caption{\label{fig:Quantum-circuit-ghz-like}Quantum circuit for ECP of non-maximally
entangled GHZ-like state $|\psi\rangle_{{\rm GHZ-like}}.$}
\end{figure}

A maximally entangled GHZ-like state is defined as \begin{equation}
\frac{|\psi_{i}0\rangle\pm|\psi_{j}1\rangle}{\sqrt{2}},\label{eq:GHZ-like-gen}\end{equation}
where $|\psi_{i}\rangle,|\psi_{j}\rangle\in\{|\psi^{\pm}\rangle,|\phi^{\pm}\rangle\}$
and $i\neq j.$ For example, we may consider a specific GHZ-like state
as \begin{equation}
|\psi\rangle=\frac{|\psi^{+}0\rangle+|\phi^{+}1\rangle}{\sqrt{2}}.\label{eq:GHZ-like-specific}\end{equation}
 Corresponding non-maximally entangled 3-qubit GHZ-like state should
be defined as

\begin{equation}
|\psi\rangle_{{\rm GHZ-like}}=\alpha|\psi^{+}0\rangle+\beta|\phi^{+}1\rangle.\label{eq:GHZ-like state}\end{equation}
As several applications of maximally entangled GHZ-like states have
been reported in recent past \cite{GHZ-like-1,GHZ-like2,GHZ-like3},
successful implementation of these applications using GHZ-like state
would require an ECP for GHZ-like state. Unfortunately no ECP for
GHZ-like state has been proposed until now. Keeping this in mind,
we wish to show that a slightly modified version of the ECP described
above for cat state works for GHZ-like states. If we start with our
defected Bell state generator as before, then with the help of an
EPR circuit (a Hadamard gate followed by a ${\rm CNOT}$ gate) and
an auxiliary qubit as shown in the left box of Fig. \ref{fig:Quantum-circuit-ghz-like}
 we can easily produce $|\psi\rangle_{{\rm GHZ-like}}$ state. However,
such construction of GHZ-like state is not an essential part of the
ECP as discussed above. Now after combining the non-maximally entangled
GHZ-like state produced as the output of the first block with a non-maximally
entangled Bell-type state $|\psi\rangle_{{\rm Bell}}$ we obtain 
\begin{widetext}
\begin{equation}
\begin{array}{lcl}
|\psi_{3}\rangle=|\psi\rangle_{{\rm Bell}}\otimes|\psi\rangle_{{\rm GHZ-like}} & = & (\alpha|00\rangle+\beta|11\rangle)_{12}\otimes(\alpha|\psi^{+}0\rangle+\beta|\phi^{+}1\rangle)_{345}\\
 & = & (\alpha^{2}|00\psi^{+}0\rangle+\alpha\beta|00\phi^{+}1\rangle+\alpha\beta|11\psi^{+}0\rangle+\beta^{2}|11\phi^{+}1\rangle)_{12345},\end{array}\label{eq:Intermidiate GHZ-like state}\end{equation}
which can be decomposed as

\begin{equation}
\begin{array}{lcl}
|\psi_{3}\rangle & = & \frac{1}{\sqrt{2}}\left[(\alpha^{2}(|\psi^{+}\rangle+|\psi^{-}\rangle)_{15}|0\psi^{+}\rangle_{234}+\alpha\beta(|\phi^{+}\rangle+|\phi^{-}\rangle)_{15}|0\phi^{+}\rangle_{234}\right.\\
 & + & \left.\alpha\beta(|\phi^{+}\rangle-|\phi^{-}\rangle)_{15}|1\psi^{+}\rangle_{234}+\beta^{2}(|\psi^{+}\rangle-|\psi^{-}\rangle)_{15}|1\phi^{+}\rangle)_{234}\right]\\
 & = & |\psi^{+}\rangle_{15}(\frac{\alpha^{2}(|000\rangle+|011\rangle)+\beta^{2}(|101\rangle+|110\rangle)_{234}}{\sqrt{2}})+|\psi^{-}\rangle_{15}(\frac{\alpha^{2}(|000\rangle+|011\rangle)-\beta^{2}(|101\rangle+|110\rangle)_{234}}{\sqrt{2}})\\
 & + & |\phi^{+}\rangle{}_{15}(\frac{\alpha\beta(|001\rangle+|010\rangle+|100\rangle+|111\rangle)_{234}}{\sqrt{2}})+|\phi^{-}\rangle_{15}(\frac{\alpha\beta(|001\rangle+|010\rangle-|100\rangle-|111\rangle)_{234}}{\sqrt{2}})\\
 & = & |\psi^{+}\rangle_{15}(\frac{\alpha^{2}(|000\rangle+|011\rangle)+\beta^{2}(|101\rangle+|110\rangle)_{234}}{\sqrt{2}})+|\psi^{-}\rangle_{15}(\frac{\alpha^{2}(|000\rangle+|011\rangle)-\beta^{2}(|101\rangle+|110\rangle)_{234}}{\sqrt{2}})\\
 & + & |\phi^{+}\rangle{}_{15}(\frac{\alpha\beta(|\psi^{+}1\rangle+|\phi^{+}0\rangle)_{234}}{\sqrt{2}})+|\phi^{-}\rangle_{15}(\frac{\alpha\beta(|\psi^{-}1\rangle+|\phi^{-}0\rangle)_{234}}{\sqrt{2}}).\end{array}\label{eq:GHZ-like state final}\end{equation}

\end{widetext}

Thus after re-ordering the qubit sequence as $12345\rightarrow15234$%
\footnote{This swapping operation that changes the particle sequence as $12345\rightarrow15234$
can be viewed as a sequence of three conventional SWAP gates that
work as follows: $12345\rightarrow15342\rightarrow15243\rightarrow15234$.%
} and a Bell measurement on first two qubits of $|\psi_{3}\rangle$
(as shown in Fig. \ref{fig:Quantum-circuit-ghz-like}) the state of
the remaining 3 qubits would collapse to one of the four states as
shown in Column 2 of Table \ref{tab:Tab2}. If the outcome of Bell
measurement is $|\psi^{\pm}\rangle$ then the protocol fails, otherwise
depending upon the outcome we apply a unitary operation as shown in
Column 3 of Table \ref{tab:Tab2}. It is easy to observe that on application
of the unitary operation in both the cases (i.e., if the outcome of
Bell measurement is $|\phi^{+}\rangle$ or $|\phi^{-}\rangle$) we
obtain a maximally entangled 3-qubit GHZ-like state. Thus we have
an ECP for GHZ-like state and as far as the main ECP part is concerned
this ECP is similar to the ECP designed for cat states with only difference
in the choice of qubits to be swapped and to be modified through unitary
operation. Thus a quantum circuit designed for ECP of cat states as
shown above will also work for GHZ-like states if suitably modified. 

\begin{widetext}
\begin{table}
\begin{centering}
\begin{tabular}{|>{\centering}p{1.5in}|>{\centering}p{2in}|>{\centering}p{1.5in}|>{\centering}p{1in}|}
\hline 
Outcome of Bell measurement on qubits $1$ and $5$ & Qubits $2,3$ and $4$ collapses to & Operation applied on qubits $2$ and $3$  & Final state of qubits $2,3$ and $4$ \tabularnewline
\hline
$|\psi^{+}\rangle_{15}$ & $\frac{\alpha^{2}(|000\rangle+|011\rangle)_{234}+\beta^{2}(|101\rangle+|110\rangle)_{234}}{\sqrt{2\left(\alpha^{4}+\beta^{4}\right)}}$ & \multicolumn{2}{c|}{Protocol fails}\tabularnewline
\hline
$|\psi^{-}\rangle_{15}$ & $\frac{\alpha^{2}(|000\rangle+|011\rangle)_{234}-\beta^{2}(|101\rangle+|110\rangle)_{234}}{\sqrt{2\left(\alpha^{4}+\beta^{4}\right)}}$ & \multicolumn{2}{c|}{Protocol fails}\tabularnewline
\hline 
$|\phi^{+}\rangle_{15}$ & $\frac{\alpha\beta(|\psi^{+}1\rangle+|\phi^{+}0\rangle)_{234}}{\sqrt{2}}$ & $I\otimes X$ & $\frac{(|\psi^{+}0\rangle+|\phi^{+}1\rangle)_{234}}{\sqrt{2}}$\tabularnewline
\hline 
$|\phi^{-}\rangle_{15}$ & $\frac{\alpha\beta(|\psi^{-}1\rangle+|\phi^{-}0\rangle)_{234}}{\sqrt{2}}$ & $iY\otimes I$ & $\frac{(|\psi^{+}0\rangle+|\phi^{+}1\rangle)_{234}}{\sqrt{2}}$\tabularnewline
\hline
\end{tabular}
\par\end{centering}

\caption{\label{tab:Tab2}Relation among outcome of Bell measurement on qubits
$1$ and $5$ and operation to be applied on the qubits $2,\,3$ to
obtain a maximally entangled GHZ-like state from a non-maximally entangled
GHZ-like state.}
\end{table}
\end{widetext}

\section{ECPs for quantum states of the form $\alpha|\Psi_{0}\rangle|0\rangle+\beta|\Psi_{1}\rangle|1\rangle$
\label{sec:A-generalized-scheme}}

In this section we will propose two ECPs for non-maximally entangled
$(n+1)$-qubit state of the following form \begin{equation}
|\psi\rangle=\alpha|\Psi_{0}\rangle|0\rangle+\beta|\Psi_{1}\rangle|1\rangle,\label{eq:channel}\end{equation}
$|\Psi_{0}\rangle$ and $|\Psi_{1}\rangle$ are arbitrary $n$-qubit
states that are mutually orthogonal. Before we propose the ECP it
would be apt to mention a few words about the relevance of the states
of the form $\alpha|\Psi_{0}\rangle|0\rangle+\beta|\Psi_{1}\rangle|1\rangle.$
This would justify why are we interested  in constructing ECPs for
states of this particular form. Clearly GHZ-like, GHZ, Bell and cat
states described above are of this form. Further, recently we have
shown that states of this particular form are useful in bidirectional
quantum teleportation \cite{bi-directional} and hierarchical quantum
communication schemes (e.g., hierarchical quantum information splitting
(HQIS), probabilistic HQIS and hierarchical quantum secret sharing
(HQSS)) \cite{HQIS}. These recently reported applications and the
fact that many well-known entangled states are of these form indicate
that the states of the form (\ref{eq:channel}) are of particular
importance. Its relevance can be further established by showing that
9 different families of SLOCC-nonequivalent 4-qubit entangled states
can be expressed in this form.

In Section \ref{sec:Introduction}, we have already mentioned that
there exist 9 families of 4-qubit entangled states. Following Verstraete
et al. \cite{4-qubit SLOCC} we may describe them as
\begin{widetext}
\[
\begin{array}{lcl}
G_{abcd} & = & \frac{a+d}{2}(|0000\rangle+|1111\rangle)+\frac{a-d}{2}(|0011\rangle+|1100\rangle)+\frac{b+c}{2}(|0101\rangle+|1010\rangle)\\
 & + & \frac{b-c}{2}(|0110\rangle+|1001\rangle),\\
L_{abc_{2}} & = & \frac{a+b}{2}(|0000\rangle+|1111\rangle)+\frac{a-b}{2}(|0011\rangle+|1100\rangle)+c(|0101\rangle+|1010\rangle)\\
 & + & |0110\rangle,\\
L_{a_{2}b_{2}} & = & a(|0000\rangle+|1111\rangle)+b(|0101\rangle+|1010\rangle)+|0110\rangle+|0011\rangle,\\
L_{ab_{3}} & = & a(|0000\rangle+|1111\rangle)+\frac{a+b}{2}(|0101\rangle+|1010\rangle)+\frac{a-b}{2}(|0110\rangle+|1001\rangle)\\
 & + & \frac{i}{\sqrt{2}}(|0001\rangle+|0010\rangle+|0111\rangle+|1011\rangle),\\
L_{a_{4}} & = & a(|0000\rangle+|0101\rangle+|1010\rangle+|1111\rangle)+(i|0001\rangle+|0110\rangle-i|1011\rangle),\\
L_{a_{2}0_{3\oplus\bar{1}}} & = & a(|0000\rangle+|1111\rangle)+|0011\rangle+|0101\rangle+|0110\rangle,\\
L_{0_{5\oplus\bar{3}}} & = & |0000\rangle+|0101\rangle+|1000\rangle+|1110\rangle,\\
L_{0_{7\oplus\bar{1}}} & = & |0000\rangle+|1011\rangle+|1101\rangle+|1110\rangle,\\
L_{0_{3\oplus\bar{1}}0_{3\oplus\bar{1}}} & = & |0000\rangle+|0111\rangle.\end{array}\]
\end{widetext}
Clearly, among 9 families listed above first 6 families (i.e., $G_{abcd},L_{abc_{2}},L_{a_{2}b_{2}},L_{ab_{3}},L_{a_{4}},L_{a_{2}0_{3\oplus\bar{1}}}$)
are parameter-dependent and remaining 3 families are parameter independent.
Now we  may note that for the specific choices of parameters $a,\, b,\, c$
and $d,$ the parameter dependent families yield different quantum
states of the form (\ref{eq:channel}) and all the parameter independent
families are already in form (\ref{eq:channel}) as shown in last
three rows of Table \ref{tab:9 families}. Specific examples of interesting
quantum states of the form (\ref{eq:channel}) obtained from the parameter
dependent families are also shown in Table \ref{tab:9 families}.
Interestingly, each of the 9 families contains state of the form (\ref{eq:channel}).
Since a state of a family can be transformed to any other state of
the family by SLOCC, so if we can construct an ECP for the quantum
states of the form (\ref{eq:channel}) in general, that would imply
that ECPs can be constructed for a large class of entangled states
involving 4-qubits. Here we further note that several applications
of the quantum states obtained as examples in $5^{th}$ Column of
Table \ref{tab:9 families} are known. For example, applications of
Bell state and GHZ state are well-known, recently protocol of quantum
dialogue using $Q_{4}$ and $Q_{5}$ is shown by us in Ref. \cite{GHZ-like2}.
Further, we have recently shown that sates of this form are useful
for various kind of hierarchical quantum communication \cite{HQIS}.
The general nature and applicability of quantum states of the form
(\ref{eq:channel}) motivated us to construct an ECP for $\alpha|\Psi_{0}\rangle|0\rangle+\beta|\Psi_{1}\rangle|1\rangle$
in general. The same is described in the following section.

\begin{widetext}
\begin{table}
\centering{}\begin{tabular}{|>{\centering}p{1.3cm}|>{\centering}p{3.3cm}|>{\centering}p{5.4cm}|>{\centering}p{6.2cm}|>{\centering}p{1.8cm}|}
\hline 
Family of states & Values of the parameters $a,b,c,d$ & Corresponding normalized states $|\psi\rangle$ & State of the form $\frac{1}{\sqrt{2}}\left(|\Psi_{0}\rangle|0\rangle+|\Psi_{1}\rangle|1\rangle\right)$
that belong to the family & Name of the state\tabularnewline
\hline 
$G_{abcd}$ & $\begin{array}{l}
a=d=\frac{1}{\sqrt{2}},b=c=0\\
a=1,b=c=d=0\end{array}$  & $\frac{1}{\sqrt{2}}(|0000\rangle+|1111\rangle),$

$\frac{1}{2}\left(|0000\rangle+|1111\rangle+|0011\rangle+|1100\rangle\right)$

$=\frac{1}{2}[(|00\rangle+|11\rangle)\otimes(|00\rangle+|11\rangle)]$ & $\frac{1}{\sqrt{2}}(|000\rangle|0\rangle+|111\rangle|1\rangle),$

$\frac{1}{\sqrt{2}}[(|0\rangle|0\rangle+|1\rangle|1\rangle]$ & cat state,

Bell state\tabularnewline
\hline 
$L_{abc_{2}}$ & $a=b=1,$ $c=0$ & $\frac{1}{\sqrt{3}}\left(|0000\rangle+|1111\rangle+|0110\rangle\right)$ & $\frac{1}{\sqrt{3}}[(|000\rangle+|011\rangle)|0\rangle+|111\rangle|1\rangle]$ & -\tabularnewline
\hline 
$L_{a_{2}b_{2}}$ & $a=1,b=0$ & $\frac{1}{2}\left(|0000\rangle+|1111\rangle+|0110\rangle+|0011\rangle\right)$ & $\frac{1}{2}[(|000\rangle+|011\rangle)|0\rangle+(|111\rangle+|001\rangle)|1\rangle]$ & -\tabularnewline
\hline 
$L_{ab_{3}}$ & $a=b=0$ & $\frac{1}{2}(|0001\rangle+|0010\rangle+|0111\rangle+|1011\rangle)$ & $\frac{1}{2}[|001\rangle|0\rangle+(|000\rangle+|011\rangle+|101\rangle)|1\rangle]$ & -\tabularnewline
\hline 
$L_{a_{4}}$ & $a=0$ & $\frac{1}{\sqrt{3}}(|0001\rangle+|0110\rangle+|1000\rangle)$ & $\frac{1}{\sqrt{3}}[(|011\rangle+|100\rangle)|0\rangle+|000\rangle|1\rangle]$ & -\tabularnewline
\hline 
$L_{a_{2}0_{3\oplus\bar{1}}}$ & $a=0$ & $|0011\rangle+|0101\rangle+|0110\rangle$$=|0\rangle\otimes\frac{1}{\sqrt{3}}(|011\rangle+|101\rangle+|110\rangle)$ & $\frac{1}{\sqrt{3}}[|11\rangle|0\rangle+(|01\rangle+|10\rangle)|1\rangle]$ & 3-qubit$W$ state\tabularnewline
\hline 
$L_{0_{5\oplus\bar{3}}}$ & parameter independent & $\frac{1}{2}\left(|0000\rangle+|0101\rangle+|1000\rangle+|1110\rangle\right)$ & $\frac{1}{2}[(|000\rangle+|100\rangle+|111\rangle)|0\rangle+|010\rangle|1\rangle]$ & $Q_{4}$ state \cite{Pati}\tabularnewline
\hline 
$L_{0_{7\oplus\bar{1}}}$ & parameter independent & $\frac{1}{2}(|0000\rangle+|1011\rangle+|1101\rangle+|1110\rangle)$ & $\frac{1}{2}[(|000\rangle+|111\rangle)|0\rangle+(|101\rangle+|110\rangle)|1\rangle]$ & $Q_{5}$ state \cite{Pati}\tabularnewline
\hline 
$L_{0_{3\oplus\bar{1}}0_{3\oplus\bar{1}}}$ & parameter independent & $|0000\rangle+|0111\rangle$$=|0\rangle\otimes\frac{1}{\sqrt{2}}(|000\rangle+|111\rangle)$ & $\frac{1}{\sqrt{2}}(|00\rangle|0\rangle+|11\rangle|1\rangle)$ & $GHZ$ state\tabularnewline
\hline
\end{tabular}\caption{\label{tab:9 families}States corresponding to the 9 families of 4-qubit
entangled states. In case of $L_{a_{4}}$ after substituting $a=0$
local unitary operations are applied to obtain the state $\frac{1}{\sqrt{3}}(|0001\rangle+|0110\rangle+|1000\rangle)$
which belongs to $L_{a_{4}}$ and suitable for the present investigation.}
\end{table}
\end{widetext}

\subsection{ECP1 for quantum states of the form \textmd{\normalsize $\alpha|\Psi_{0}\rangle|0\rangle+\beta|\Psi_{1}\rangle|1\rangle$ }}

\begin{figure}
\begin{centering}
\includegraphics[scale=0.67]{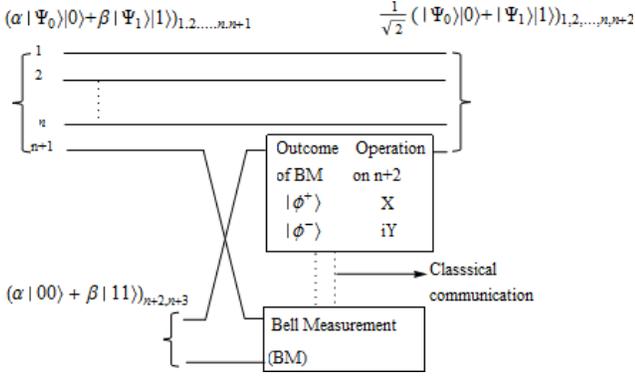}
\par\end{centering}

\caption{\label{fig:General-circuit-for}Quantum circuit for ECP for partially
entangled $(n+1)$-qubit state of the form $\alpha|\Psi_{0}\rangle|0\rangle+\beta|\Psi_{1}\rangle|1\rangle.$
}
\end{figure}

The ECP is described here through the quantum circuit shown in the
Fig. \ref{fig:General-circuit-for}. In this quantum circuit initial
$(n+1)$-qubit state $\alpha|\Psi_{0}\rangle|0\rangle+\beta|\Psi_{1}\rangle|1\rangle$
is combined with $|\psi\rangle_{{\rm Bell}}$ and we obtain the combined
input state as 
\begin{widetext}
\[
\begin{array}{lcl}
|\psi_{5}\rangle & = & \left(\alpha|\Psi_{0}\rangle|0\rangle+\beta|\Psi_{1}\rangle|1\rangle\right)_{1,2,\cdots,n+1}\otimes(\alpha|00\rangle+\beta|11\rangle)_{n+2,n+3}\\
 & = & \left(\alpha^{2}|\Psi_{0}\rangle|000\rangle+\alpha\beta|\Psi_{1}\rangle|100\rangle+\alpha\beta|\Psi_{0}\rangle|011\rangle+\beta^{2}|\Psi_{1}\rangle|111\rangle\right)_{1,2,\cdots,n+1,n+2,n+3}.\end{array}\]
After swapping the $(n+1)$-th with $(n+2)$-th qubits we obtain \[
\begin{array}{lcl}
|\psi_{6}\rangle & = & \left(\alpha^{2}|\Psi_{0}\rangle|000\rangle+\alpha\beta|\Psi_{1}\rangle|010\rangle+\alpha\beta|\Psi_{0}\rangle|101\rangle+\beta^{2}|\Psi_{1}\rangle|111\rangle\right)_{1,2,\cdots,n,n+2,n+1,n+3}\\
 & = & \frac{1}{\sqrt{2}}\left(\left(\alpha^{2}|\Psi_{0}\rangle|0\rangle+\beta^{2}|\Psi_{1}\rangle|1\rangle\right)|\psi^{+}\rangle+\left(\alpha^{2}|\Psi_{0}\rangle|0\rangle-\beta^{2}|\Psi_{1}\rangle|1\rangle\right)|\psi^{-}\rangle\right.\\
 & + & \left.\alpha\beta\left(|\Psi_{0}\rangle|1\rangle+|\Psi_{1}\rangle|0\rangle\right)|\phi^{+}\rangle+\alpha\beta\left(|\Psi_{0}\rangle|1\rangle-|\Psi_{1}\rangle|0\rangle\right)|\phi^{-}\rangle\right)_{1,2,\cdots,n,n+2,n+1,n+3}.\end{array}\]
\end{widetext}

Now a Bell measurement is performed on the last two qubits of $|\psi_{6}\rangle.$
If the Bell measurement yields $|\psi^{\pm}\rangle$ then the protocol
fails, but if it yields $|\phi^{+}\rangle$ $\left(|\phi^{-}\rangle\right)$
then we can obtain the desired state (i.e., $\left(\frac{|\Psi_{0}\rangle|0\rangle+|\Psi_{1}\rangle|1\rangle}{\sqrt{2}}\right)_{1,2,\cdots,n,n+2}$)
by applying $X$ $(iY)$ on the $(n+2)$-th qubit. This provides a
simple, but very useful ECP schemes for quantum states of the form
(\ref{eq:channel}) in general. Consequently, we obtain ECP for GHZ-like
state, GHZ-state, 9 families of SLOCC-nonequivalent 4-qubit entangled
state, cluster state, cat state etc. In this ECP, knowledge of $\alpha,$
$\beta$ is required for the construction of $|\psi\rangle_{{\rm Bell}},$
but $\alpha,\,\beta$ can be complex. In what follows we propose another
alternative quantum circuit for ECP of $\alpha|\Psi_{0}\rangle|0\rangle+\beta|\Psi_{1}\rangle|1\rangle.$ 

\subsection{ECP2 for quantum states of the form \textmd{\normalsize $\alpha|\Psi_{0}\rangle|0\rangle+\beta|\Psi_{1}\rangle|1\rangle$ }}

\begin{figure}
\begin{centering}
\includegraphics[scale=0.67]{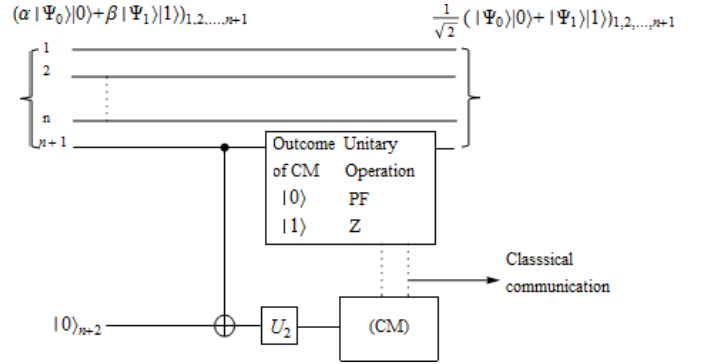}
\par\end{centering}

\caption{\label{fig:alternative ckt for ECP} An alternative quantum circuit
for ECP for partially entangled $(n+1)$-qubit state of the form $\alpha|\Psi_{0}\rangle|0\rangle+\beta|\Psi_{1}\rangle|1\rangle$.
Here CM stands for measurement in computational basis and PF stands
for protocol fails.}
\end{figure}

In this subsection we propose an alternative circuit for ECP for quantum
states of the form $\alpha|\Psi_{0}\rangle|0\rangle+\beta|\Psi_{1}\rangle|1\rangle$.
The proposed ECP is shown in the Fig. \ref{fig:alternative ckt for ECP}.
In this case input state is \[
\begin{array}{lcl}
|\psi_{5}\rangle & = & \left(\alpha|\Psi_{0}\rangle|0\rangle+\beta|\Psi_{1}\rangle|1\rangle\right)_{1,2,\cdots,n+1,}\otimes|0\rangle_{n+2}\\
 & = & \frac{1}{\sqrt{2}}\left(\alpha|\Psi_{0}\rangle|00\rangle+\beta|\Psi_{1}\rangle|10\rangle\right)_{1,2,\cdots,n+1,n+2,}\end{array}\]
 Now a ${\rm CNOT}_{\left(n+1\right)\rightarrow\left(n+2\right)}$
gate is applied on $|\psi_{5}\rangle$ using $(n+1)$-th qubit as
control qubit and $(n+2)$-th qubit as target qubit to yield \[
|\psi_{6}\rangle=\frac{1}{\sqrt{2}}\left(\alpha|\Psi_{0}\rangle|00\rangle+\beta|\Psi_{1}\rangle|11\rangle\right)_{1,2,\cdots,n+1,n+2}.\]
 Now we may apply a unitary operator $U_{2}=\left(\begin{array}{cc}
\alpha & \beta\\
-\beta & \alpha\end{array}\right)$ on $(n+2)$-th qubit to obtain (here $\alpha,$ $\beta$ are real)%
\footnote{Here $U_{2,n+2}|\psi_{6}\rangle$ denotes that single qubit operation
$U_{2}$ operates on $(n+2)$-th qubit of $|\psi_{6}\rangle$ and
identity operators operate on rest of the qubits.%
}
\begin{widetext}
\[
\begin{array}{lcl}
|\psi_{7}\rangle & = & U_{2,n+2}|\psi_{6}\rangle\\
 & = & \left(\alpha^{2}|\Psi_{0}\rangle|00\rangle-\alpha\beta|\Psi_{0}\rangle|01\rangle+\beta^{2}|\Psi_{1}\rangle|10\rangle+\alpha\beta|\Psi_{1}\rangle|11\rangle\right)_{1,2,\cdots,n+1,n+2}\\
 & = & \left(\left(\alpha^{2}|\Psi_{0}\rangle|0\rangle+\beta^{2}|\Psi_{1}\rangle|1\rangle\right)|0\rangle-\alpha\beta\left(|\Psi_{0}\rangle|0\rangle-|\Psi_{1}\rangle|1\rangle\right)|1\rangle\right)_{1,2,\cdots,n+1,n+2},\end{array}\]
\end{widetext}
and subsequently measure the last qubit (i.e., $\left(n+2\right)$-th
qubit) in computational basis. If the measurement result yields $|0\rangle$
then the protocol fails, otherwise we apply $Z=\sigma_{z}$ gate on
the $\left(n+1\right)$-th qubit to obtain the desired maximally entangled
state: \[
\left(\frac{|\Psi_{0}\rangle|0\rangle+|\Psi_{1}\rangle|1\rangle}{\sqrt{2}}\right)_{1,2,\cdots,n+1}.\]
Thus we have two alternative ECPs for quantum states of the form (\ref{eq:channel})
in general. In the first scheme (ECP1) we don't need $\alpha,\beta$
to be real, while the same is required in the second scheme (ECP2).
This is indicative of superiority of ECP1 over ECP2. Still ECP2 is
interesting for its extreme simplicity. Specifically, ECP2 neither
require any Bell measurement nor any particle swapping. Further, as
the amount of initial entanglement is less compared to ECP1, its efficiency
would be more compared to ECP1 if we use Sheng et al.'s quantitative
measure of quality of ECP \cite{Sheng-1}. This point is elaborated
in Section \ref{sec:Efficiency}.

\section{Optical implementation \label{sec:Optical-implementation}}

\begin{figure}
\begin{centering}
\includegraphics[scale=0.8]{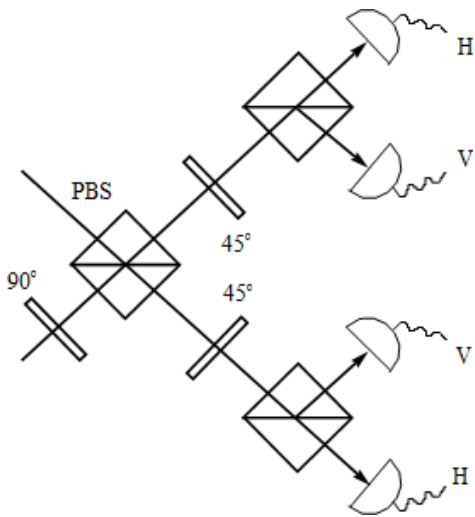}
\par\end{centering}

\caption{\label{fig:Bell measurement}A linear optics based scheme for Bell-state
measurement for (a) single photon polarization qubit. The scheme uses
polarizing Beam splitters (PBSs) that allows horizontally polarized
photon to transmit and reflects the vertically polarized photon, wave
plates, and on/off photo detectors \cite{Lee-Bell measuremnt}.}
\end{figure}

Bell measurement was used in one of the pioneering work on ECP \cite{S. Bose }.
However, in most of the subsequent works the use of Bell measurement
was criticized by noting that the Bell measurement is a complex task
and using linear optics we cannot distinguish all the four Bell states
\cite{criticized bell}. Here it is important to note that all the
Bell-measurement based ECPs proposed here fail when outcome of the
Bell measurement is $|\psi^{\pm}\rangle$. Thus for the practical
implementation of the first two ECPs proposed here it would be sufficient
to distinguish $|\phi^{+}\rangle$ and $|\phi^{-}\rangle.$ A simple
linear optics setup shown in Fig. \ref{fig:Bell measurement} can
perform the task \cite{Lee-Bell measuremnt}. The working of the optical
circuit shown in Fig. \ref{fig:Bell measurement} is elaborately described
in Ref. \cite{Lee-Bell measuremnt}. Here for the completeness of
our discussion we may briefly note that when information is encoded
using polarization degree of freedom then usually horizontal ($H$)
and vertical ($V$) polarized states represent $|0\rangle$ and $|1\rangle,$
respectively. Thus Bell states can be expressed as $|\psi^{\pm}\rangle=\frac{1}{\sqrt{2}}\left(|HH\rangle\pm|VV\rangle\right)$
and $|\phi^{\pm}\rangle=\frac{1}{\sqrt{2}}\left(|HV\rangle\pm|VH\rangle\right)$.
If $|\phi^{+}\rangle$ enters the optical circuit then the detectors
click as either $H_{{\rm up}},V_{{\rm down}}$ or $V_{{\rm {\rm up}}},H_{{\rm down}}$
where the subscript up (down) denotes the outcome of top (bottom)
two detectors. Similarly, when $|\phi^{-}\rangle$ enters the optical
circuit then the detectors clicks as either $H_{{\rm up}},H_{{\rm down}}$
or $V_{{\rm {\rm up}}},V_{{\rm down}}$. Thus $|\phi^{+}\rangle$
can be distinguished from $|\phi^{-}\rangle$. However, we cannot
distinguish $|\psi^{+}\rangle$ and $|\psi^{-}\rangle$ as in both
of the cases detectors click as $H_{{\rm up}},V_{{\rm up}}$ or $H_{{\rm {\rm down}}},V_{{\rm down}}$.
Thus for the Bell-measurement based ECPs proposed here if both of
the upper or lower detectors click then the protocol fails, otherwise
we apply appropriate unitary operations as described above. Further,
the ${\rm CNOT}$ used in ECP2 can be implemented using optical circuits
implemented by J. L. O\textquoteright{}Brien et al. \cite{CNOT-obrien}.
Thus in general ECPs proposed here can be realized optically. However,
the applicability of the circuits is not limited to optical realization.
For example, these ECPs may be practically realized using NMR as Bell
measurement is possible in NMR based technologies \cite{NMR-Anil Kumar}.

\section{Efficiency\label{sec:Efficiency}}

\begin{figure}
\begin{centering}
\includegraphics[scale=0.4]{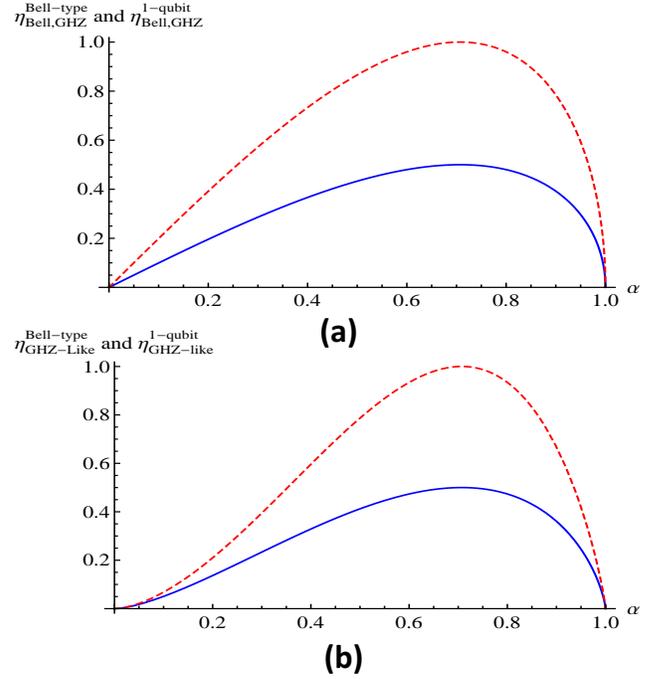}
\par\end{centering}

\caption{\label{fig:Variaontion-of-eta}(Color online) Variation of $\eta$ with $\alpha$.
(a) for Bell and GHZ states and (b) for GHZ-like state. Solid line
represents ECPs realized using assistance of $|\psi\rangle_{{\rm Bell}}$
and the dashed line represent single-qubit assisted ECP. }
\end{figure}

Recently Sheng et al. \cite{Sheng-1} have introduced a quantitative
measure of entanglement concentration efficiency of an ECP. They have
referred to it as \emph{entanglement transformation efficiency} $\eta$
and explicitly defined it as

\begin{equation}
\eta=\frac{E_{c}}{E_{0}}\label{eq:efficiency1}\end{equation}
where, $E_{0}$ is the amount of entanglement in the initial partially
entangled state and $E_{c}$ is the amount of entanglement of the
state after concentration. Further, they have defined $E_{c}$ as

\begin{equation}
E_{c}=P_{s}\times E_{m}+(1-P_{s})E^{\prime}\label{eq:efficiency2}\end{equation}
where $P_{s}$ is the success probability of obtaining the maximally
entangled state on execution of the ECP and $E_{m}$ is the amount
of entanglement in the maximally entangled state. Sheng et al. assumed
that the measure of entanglement is chosen in such a way that the
amount of entanglement in a maximally entangled state is 1, however
it may be different in general. For example, we often use a definition
of negativity where negativity of maximally entangled Bell state is
$0.5$ and log negativity is $1.$ Thus the first (second) term of
$E_{c}$ corresponds to success (failure) of the ECP. Up to this point
the definition of $\eta$ seems clear and straight forward. However,
there exists an ambiguity in the definition, it is not clear whether
$E_{0}$ is the amount of entanglement of the state to be concentrated
or that of the entire initial state. To remove this ambiguity we choose
$E_{0}$ to be the total initial entanglement. This choice naturally
implies higher efficiency of single photon assisted ECPs over Bell-type
state assisted ECPs (such as the first 2 ECPs of the present paper).
Further, a closer look into (\ref{eq:efficiency1}) would reveal that
it is neither unique nor easily expendable to the multipartite case.
Specifically, the definition does not define which measure of entanglement
is to be used for obtaining $E_{0}.$ Existence of different measures
of bipartite entanglement and the fact that these measures are not
monotone of each other makes the definition (\ref{eq:efficiency1})
non-unique. In fact, Sheng et al. \cite{Sheng-1} have used von Neumann
entropy as a measure of entanglement, but von Neumann entropy is a
good measure of entanglement for bipartite systems only. This limitation
exists for most of the well known measures of entanglement and this
fact leads to an interesting question: How to find $\eta$ for an
ECP that is designed for multipartite case. Interestingly, the problem
is equivalent to provide a quantitative measure of multipartite entanglement.
In last two decades several efforts have been made to introduce measures
of multipartite entanglement \cite{ent-measure-1,ent-measure-2,ent-measure-3,ent-measure-4,ent-measure-5,ent-measure-6}.
We may use some of the approaches followed in \cite{ent-measure-1,ent-measure-2,ent-measure-3,ent-measure-4,ent-measure-5,ent-measure-6}
to obtain $\eta$ for multipartite ECP. To show the dependence of
$\eta$ one choice of entanglement measure we may choose \emph{tangle}
\cite{ent-measure-2,ent-measure-3,ent-measure-5} as a measure of
entanglement. In that case, entanglement of $|\psi\rangle_{{\rm Bell}}$
is $4|\alpha\beta|^{2}$ and success probability for all ECPs presented
here and for Sheng et al.S ECP is $2|\alpha\beta|^{2}.$ Thus $\eta$
for Sheng protocol and our last protocol (i.e., ECP2) will be $\frac{1}{2}.$
Whereas that of our first two ECPs and ECP of Zhao \cite{zhao-proposal}
will be $\frac{1}{4}.$ As the tangle for a partially entangled GHZ
state is same as that of a partially entangled Bell state. Efficiency
of our protocol would remain same ($\frac{1}{4}$) for ECP for partially
entangled GHZ state. Clearly in all these three cases $\eta$ is independent
of $\alpha$ which is in contrast with the result obtained by Sheng
using von Neumann entropy as a measure of entanglement. Specifically,
they had observed $\eta$ was a function of $\alpha$. To extend the
definition of efficiency $\eta$ to the multipartite case and to elaborate
it's dependence on choice of the entanglement measure we may note
that in 2004, Yu and Song established \cite{ent-measure-4} that any
good measure $M_{A-B}$ of bi-partite entanglement
can be generalized to multipartite systems, by considering bipartite
partitions of the multipartite system. Yu and Song defined a simple
measure of tripartite entanglement as \begin{equation}
M_{ABC}=\frac{1}{3}\left(M_{A-BC}+M_{B-AC}+M_{C-AB}\right),\label{eq:yu song}\end{equation}
where $M_{i-jk}$ is a measure of entanglement between subsystem $i$
and subsystem $jk$. $M_{i-jk}$ may be any good measure of bi-partite
entanglement (e.g., von Neumann entropy, negativity etc.). Yu and
Song's idea was used to measure tripartite entanglement in various
systems using different measures of bipartite entanglement e.g., negativity,
concurrence, and von Neumann\textquoteright{}s entropy (cf. \cite{ent-measure-1}
and references therein). However, some limitations of the above measure
(\ref{eq:yu song}) were found and a new measure of tripartite entanglement
was introduced by Sab\i{}n and Garca-Alcaine by replacing arithmetic
mean present in (\ref{eq:yu song}) by geometric mean. Thus Sab\i{}n
and Garca-Alcaine's measure of tripartite entanglement is given as
\cite{ent-measure-1}\begin{equation}
M_{ABC}=\left(M_{A-BC}M_{B-AC}M_{C-AB}\right)^{\frac{1}{3}}.\label{eq:sabin}\end{equation}
In what follows we have provided analytic expressions for efficiencies
of ECPs proposed here using (\ref{eq:efficiency1}) and (\ref{eq:sabin}).
To be precise, if we use negativity as a measure of bipartite entanglement
then the efficiency of the first two protocols (i.e., protocols assisted
by $|\psi\rangle_{{\rm Bell}}$) proposed here are as follows \begin{equation}
\eta_{{\rm Bell},{\rm GHZ}}^{{\rm Bell-type}}=\frac{2|\alpha\beta|^{2}}{2|\alpha\beta|}=|\alpha\beta|,\label{eq:eta1}\end{equation}
 and 

\begin{equation}
\eta_{{\rm GHZ-like}}^{{\rm Bell}{\rm -type}}=\frac{2|\alpha\beta|^{2}}{\sqrt[3]{\frac{1}{4}|\alpha\beta|}+|\alpha\beta|}.\label{eq:eta2}\end{equation}
 Similarly, in case of ECP2 (i.e., for the single qubit assisted protocol)
we obtain \begin{equation}
\eta_{{\rm Bell},{\rm GHZ}}^{{\rm 1-qubit}}=2|\alpha\beta|,\label{eq:eta3}\end{equation}
 and 

\begin{equation}
\eta_{{\rm GHZ-like}}^{{\rm 1-qubit}}=\frac{4|\alpha\beta|^{2}}{\sqrt[3]{2|\alpha\beta|}}.\label{eq:eta4}\end{equation}
From the above equations it is clear that the ECP2 is more efficient
than ECP1. The same is illustrated in Fig. \ref{fig:Variaontion-of-eta}.
In the left (right) panel of  Fig. \ref{fig:Variaontion-of-eta} the
variation of efficiency of ECPs proposed for partially entangled Bell
and GHZ state (GHZ-like state) with $\alpha$ are shown.

\section{Conclusion\label{sec:Conclusion}}

We have proposed three ECPs in the present paper. The first one is
shown to generate a maximally entangled cat state from the corresponding
partially entangled state. A modified version of this ECP is also
introduced as an ECP for GHZ-like state. ECPs for cat states were
proposed earlier, too. However, no ECP for GHZ-like states were proposed
until now. Thus this is the first ever ECP reported precisely for
GHZ-like state. The last two ECPs are designed for quantum states
of the general form $\alpha|\Psi_{0}\rangle|0\rangle+\beta|\Psi_{1}\rangle|1\rangle.$
These two protocols are extremely interesting as several applications
of the states of the form $\alpha|\Psi_{0}\rangle|0\rangle+\beta|\Psi_{1}\rangle|1\rangle$
are reported in recent past \cite{GHZ-like-1,GHZ-like2,GHZ-like3,bi-directional,HQIS}.
Further, its very important as specific states from the 9 families
of SLOCC-nonequivalent 4-qubit entangled states can be described in
this form. Thus the proposed ECPs are valid for the 9 families of
4-qubit entangled states. Further, partially entangled cat-like, GHZ-like,
GHZ, W and Bell states can also be expressed as $\alpha|\Psi_{0}\rangle|0\rangle+\beta|\Psi_{1}\rangle|1\rangle$.
The ECPs are not described in the usual style, rather they are described
as quantum circuit. From the quantum circuits described above one
can clearly see that all the ECPs proposed here require local measurement,
classical communication and post selection. According to Vedral et
al. \cite{Vedral} these are the basic steps required by any good
ECP or EP. Further, the efficiency of the proposed protocols are discussed
in detail using a quantitative measure of efficiency that was recently
introduced by Sheng et al. \cite{Sheng-1}. Apparently any Bell-type
state assisted ECP (e.g., the first two ECPs of the present paper
and the Zhao et al. proposal \cite{zhao-proposal}) will have lesser
efficiency compared to linear optics-based single qubit assisted ECPs
\cite{Sheng-1}. Again if we go beyond linear optics and use nonlinear
resources then the efficiency would increase further. However, this
parameter cannot be considered as a basis of choosing ECPs as the
measure of efficiency introduced by Sheng et al. \cite{Sheng-1} is
really a weak measure. Keeping these in mind and the fact that the
proposed ECPs that are applicable to a large class of quantum states
of practical interest are experimentally realizable using linear optical
resources and NMR, we conclude the paper with an expectation that
experimentalists will find it interesting to implement the ECPs proposed
here. 

\textbf{Acknowledgment: }AP thanks Department of Science and Technology
(DST), India for support provided through the DST project No. SR/S2/LOP-0012/2010
and he also acknowledges the supports received from the projects CZ.1.05/2.1.00/03.0058
and CZ.1.07/2.3.00/20.0017 of the Ministry of Education, Youth and
Sports of the Czech Republic.

\end{document}